\documentclass[cameraready]{Interspeech}

\title{Edit Content, Preserve Acoustics: Imperceptible Text-Based Speech Editing via Self-Consistency Rewards}

\author[affiliation={1,2}, orcid=0009-0000-9015-000X]{Yong}{Ren}
\author[affiliation={3}, orcid=0000-0003-2422-4618, correspondingauthor]{Jiangyan}{Yi}
\author[affiliation={3,4}, orcid=0000-0002-9344-6428, correspondingauthor]{Jianhua}{Tao}
\author[affiliation={1}, orcid=0000-0003-1490-6973]{Tao}{Wang}
\author[affiliation={1}, orcid=0009-0004-3410-354X]{Le}{Xu}
\author[affiliation={4}, orcid=0009-0004-1632-4293, correspondingauthor]{Zhengqi}{Wen}

\address{
    $^1$ The State Key Laboratory of Multimodal Artificial Intelligence Systems, Institute of Automation, Chinese Academy of Sciences 
    $^2$ School of Artificial Intelligence, University of Chinese Academy of Sciences 
    $^3$ Department of Automation, Tsinghua University
    $^4$ BNRist, Tsinghua University
}

\email{thurenyong@gmail.com, yijy@tsinghua.edu.cn, jhtao@tsinghua.edu.cn, zqwen@tsinghua.edu.cn}


\usepackage{comment}

\usepackage{amsmath,amssymb,amsfonts}
\usepackage{mathabx}
\usepackage{textcomp}
\usepackage{graphicx}
\usepackage{hyperref}
\usepackage[table]{xcolor}
\usepackage{comment}
\usepackage{multirow}
\usepackage{arydshln}
\usepackage{subfigure}
\usepackage{ulem}
\usepackage{algorithm}
\usepackage{algorithmic}
\usepackage{cite}
\usepackage{makecell}
\usepackage{booktabs}
\usepackage{mwe}
\usepackage{enumitem}
\setlist[itemize]{noitemsep, topsep=0pt, leftmargin=*}

\begin{document}

\maketitle

\begin{abstract}
    Imperceptible text-based speech editing modifies spoken content through transcript manipulation while preserving acoustic continuity. Prior acoustic-space approaches suffer from content–style entanglement, causing unstable generation and boundary artifacts. We introduce a framework guided by the principle `Edit Content, Preserve Acoustics'. Editing is conducted in a stable semantic space, while acoustic realization is handled by a Flow Matching decoder. To ensure perceptual consistency, we propose Self-Consistency Rewards Group Relative Policy Optimization, which leverages a pre-trained Text-to-Speech model as an implicit critic, together with intelligibility and duration constraints. Experiments demonstrate consistent improvements over state-of-the-art autoregressive and non-autoregressive baselines in intelligibility, robustness, and perceptual quality.
\end{abstract}

\section{Introduction}
\label{sec:intro}

Text-based speech editing modifies spoken content by editing transcripts, enabling word insertion, deletion, or substitution without costly re-recording~\cite{jin2017voco, tan2021editspeech, wang2022campnet,wang2022context, wang2024emotion}. The capability is critical for applications such as podcast correction, audiobook revision, and post-production dialogue editing~\cite{peng2024voicecraft, kim2025instance}.

Despite significant progress, achieving `imperceptible' editing remains challenging. Early Non-Autoregressive (NAR) speech editing methods~\cite{wang2022campnet, bai20223, wang2025speechpalette, liu2025fluenteditor2} offer stable inference but struggle with long-range dependency modeling, leading to flattened prosody. Conversely, recent Autoregressive (AR) speech editing models based on Neural Codec Language Models (NCLMs)~\cite{peng2024voicecraft,zheng2025voicecraft,mohammad2025speak,yang2023uniaudio,yan2025ming} have achieved state-of-the-art (SOTA) naturalness. However, these methods typically operate on acoustic tokens~\cite{defossez2023high,kumar2023high} where content and style are entangled. This coupling often compromises robustness, leading to hallucinations and boundary artifacts when modifying content~\cite{huang2025voicenong}.

To advance the state of text-based speech editing, we revisit the task and contend that it differs fundamentally from Text-to-Speech (TTS). We characterize text-based speech editing as a context-constrained incremental generation problem. Achieving imperceptibility requires balancing modification with continuity; to this end, we propose a framework grounded in the principle of \textit{Edit Content, Preserve Acoustics}.


\textbf{Structural Foundations for Acoustic Preservation.}
A key limitation of existing text-based speech editing approaches is their direct manipulation of acoustic representations, where linguistic content and timbre are tightly coupled. This entanglement makes joint prediction unstable and often leads to hallucinations or boundary artifacts. To address this issue, we decouple the editing process by performing content modification in a disentangled semantic space that captures linguistic content and coarse prosody. Acoustic reconstruction is subsequently handled by a Flow Matching~\cite{lipmanflow} decoder. This hierarchical design projects both edited regions and original context into a unified acoustic manifold, preserving acoustic coherence while enabling precise content manipulation.

\begin{figure*}[ht!]
  \centering
  \includegraphics[width=\linewidth]{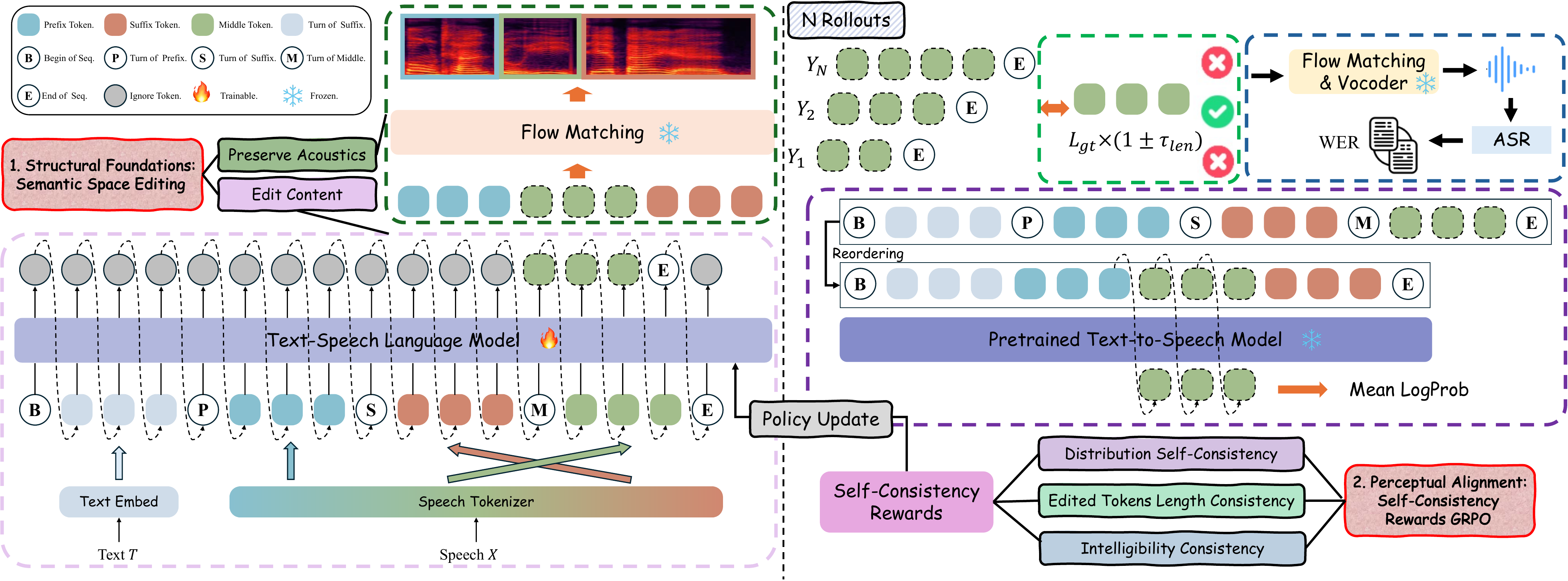}
\caption{The overall framework of our proposed method. The pipeline consists of two stages: (1) \textbf{Structural Foundations} (Left), where we employ a semantic-token-based LLM for conditional content infilling followed by a Flow Matching decoder for acoustic reconstruction; and (2) \textbf{Perceptual Alignment} (Right), where the policy is fine-tuned via Self-Consistency Rewards GRPO.}  \label{fig:architecture}
\vspace{-8pt}
\end{figure*}

\textbf{Perceptual Alignment for Further Coherence.} Although structural decoupling maintains the acoustic foundation (such as timbre and acoustic environment), semantic tokens inherently encode rhythm and paralinguistic information. Consequently, ensuring that the edited content fuses indistinguishably with the utterance requires explicit perceptual alignment. 
Reinforcement Learning (RL) has emerged as a powerful paradigm for aligning Large Language Models (LLMs)~\cite{shao2024deepseekmath, uesato2022solving, lightman2023let}, with initial explorations in speech generation~\cite{zhang2024speechalign, zhong2025multi, gao2025explore, ren2026evaluating}.
However, existing speech rewards typically target Text-to-Speech (TTS) metrics like intelligibility or speaker similarity, failing to address the seamless fusion of the edited region with the surrounding context beyond mere timbre coherence. To address this, we introduce a novel alignment mechanism using a pre-trained Text-to-Speech (TTS) model as an implicit critic. Based on the premise that a powerful TTS model captures the distribution of natural speech, we utilize the conditional likelihood of the edited tokens given the context as a statistical proxy for the coherence between the edited region and the entire sentence. Complemented by strict constraints on Automatic Speech Recognition (ASR) Word Error Rate (WER) and duration validity, we construct a composite reward that aligns the generation with human perceptual expectations.
In summary, our contributions are as follows: 
\begin{itemize} 
    \item We propose a novel framework for imperceptible text-based speech editing, addressing the problem through structural foundations and perceptual alignment. 
    \item \textbf{Semantic Space Editing.} We adopt a semantic-token-based architecture to decouple content editing from acoustic reconstruction, significantly reducing artifacts compared to acoustic-token-based baselines. 
    \item \textbf{Self-Consistency Rewards GRPO.} We are the first to utilize a pre-trained TTS model as a consistency critic within a RL framework for speech editing. Combined with ASR and length rewards, it effectively enhances global coherence and achieves perceptual alignment.
    \item Empirical evaluations demonstrate that our method significantly outperforms NAR and AR baselines, achieving superior intelligibility, robustness and perceptual quality. 
\end{itemize}

\section{Method}

This section presents the proposed framework for imperceptible text-based speech editing. We first describe the overall architecture, followed by the formulation of \textbf{Structural Foundations} and the \textbf{Perceptual Alignment} mechanism.

\subsection{Overall architecture}

Guided by the principle of \textit{Edit Content, Preserve Acoustics}, we design a hierarchical framework illustrated in Figure~\ref{fig:architecture}. The system operates in two stages:
\begin{itemize}
    \item \textbf{Structural Foundations via Decoupling:}
    Editing is restricted to a discrete semantic space to isolate linguistic manipulation from acoustic variations. The edited semantics are subsequently rendered into waveform space by a Flow Matching decoder within a unified acoustic manifold.
    
    \item \textbf{Perceptual Alignment via RL:}
    To ensure the edited region fuses indistinguishably with the bidirectional context, we introduce a \textit{Self-Consistency Rewards GRPO} stage. We leverage a pre-trained TTS model as an implicit critic to maximize the conditional likelihood of the generated tokens, thereby enhancing global coherence of semantic tokens without requiring paired ground-truth data.
\end{itemize}

\subsection{Structural Foundations: Semantic Space Editing}

Prior AR speech editing approaches operate in acoustic-token space, where linguistic content and timbre are inherently entangled~\cite{peng2024voicecraft}. To address this issue, we adopt a decoupled architecture consisting of an LLM for semantic generation and a Flow Matching decoder for acoustic reconstruction~\cite{du2025cosyvoice}.

We formulate text-based speech editing as a conditional token infilling problem in semantic space. To facilitate this, we employ a Prefix-Suffix-Middle (PSM) formatting strategy. Given an input waveform $X$, a semantic tokenizer $\mathcal{T}(\cdot)$ encodes it into a discrete token sequence $ S = \mathcal{T}(X). $
The sequence is partitioned into three segments,
$\mathbf{S}_{\text{pre}}$, $\mathbf{S}_{\text{mid}}$, and $\mathbf{S}_{\text{suf}}$,
representing the left context, editable region, and right context, respectively. The model input $\mathbf{Q}$ is then constructed by: 

\begin{equation}
\mathbf{Q} = [\mathbf{T}; \mathbf{S}_{\text{pre}}; \mathbf{S}_{\text{suf}}], 
\end{equation}
where $\mathbf{T}$ denotes the tokenized transcription.

We employ a decoder-only transformer as the policy model $\pi_\theta$. During supervised training, the model minimizes negative log-likelihood of the missing middle tokens conditioned on $\mathbf{Q}$:

\begin{equation}
    \small
    \mathcal{L}(\theta) = - \sum_{t=1}^{|\mathbf{S}_{\text{mid}}|} \log \pi_\theta(s_{\text{mid}, t} \mid \mathbf{Q}, s_{\text{mid}, <t}).
\end{equation}

This objective enables coherent semantic infilling between the prefix and suffix, after which the waveform is reconstructed using a Flow Matching decoder and vocoder.

\subsection{Perceptual Alignment: Self-Consistency Rewards GRPO}

Although semantic-space editing preserves acoustic structure, semantic tokens still encode prosodic and paralinguistic cues, causing autoregressive sampling to produce hallucinations or mismatches. To achieve imperceptible editing, we introduce a perceptual alignment mechanism that enforces statistical consistency with the surrounding context under the natural speech distribution. This is realized through a novel \textit{Self-Consistency Rewards GRPO}, illustrated in Figure~\ref{fig:architecture} (right).

GRPO estimates the baseline from the relative performance of multiple samples generated for the same prompt. For each editing query $\mathbf{Q}$, we sample a group of $G$ candidate sequences $\{\mathbf{o}_1,\dots,\mathbf{o}_G\}$ from the previous policy $\pi_{\theta_{\text{old}}}$. The optimization relies on the Relative Advantage $\hat{A}_i$, which measures each sample against its peers. Formally, given rewards $\{R_1,\dots,R_G\}$, the advantage is defined as:

\begin{equation}
\hat{A}_i =
\frac{R_i - \mu_R}{\sigma_R},
\end{equation}
where $i$ indexes the $i$-th completion among the $G$ candidates for input $\mathbf{Q}$, and $\mu_R$ and $\sigma_R$ denote the mean and standard deviation of rewards within the sampled group.

\subsubsection{Log-Probability Self-Consistency Reward ($r_{\text{sc}}$)}

We leverage a frozen pre-trained TTS model as an implicit critic that evaluates the likelihood of edited tokens under the natural speech distribution.

\textit{Formulation.}
Given a generated token sequence $\hat{\mathbf{S}}_{\text{mid}}$, the reward is defined as the average log-likelihood under the frozen TTS reference model $\pi_{\text{tts}}$:
\begin{equation}
\small
r_{\text{sc}}=
\frac{1}{|\hat{\mathbf{S}}_{\text{mid}}|}
\sum_{t=1}^{|\hat{\mathbf{S}}_{\text{mid}}|}
\log \pi_{\text{tts}}
(\hat{s}_{\text{mid},t}\mid
[\mathbf{T};\mathbf{S}_{\text{pre}}],
\hat{s}_{\text{mid},<t}).
\end{equation}

\subsubsection{Intelligibility Reward ($r_{\text{wer}}$)}

Optimizing only $r_{\text{sc}}$ may lead to reward hacking, where high-likelihood yet semantically trivial outputs (e.g., silence or repetition) are favored. To enforce content correctness, we introduce an intelligibility reward:
\begin{equation}
    \small
    r_{\text{wer}} = 1 - \operatorname{WER}(\mathbf{W}_{\text{rec}}, \mathbf{T}_{\text{tgt}}),
\end{equation}
where $\operatorname{WER}$ is computed by an ASR model on the reconstructed waveform $\mathbf{W}_{\text{rec}}$. 

\begin{table*}[h]
\centering
\caption{Performance comparison on the text-based speech editing benchmark. The symbol '$\diamond$' indicates that the results are cited directly from the original paper~\cite{yan2025ming}. \textbf{\textcolor{red}{Red}} indicates the best result, and \textbf{\textcolor{blue}{Blue}} indicates the second best.}
\label{tab:main_results_edit}
\resizebox{0.85\textwidth}{!}{%
\begin{tabular}{@{}llcccc@{}}
\toprule
\textbf{Edit Type} & \textbf{Model} & \multicolumn{4}{c}{\textbf{Performance}} \\
\midrule

& & \textbf{WER(\%)$\downarrow$ basic \textbar{} full} & \textbf{SIM$\uparrow$ basic \textbar{} full} & \textbf{DNSMOS$\uparrow$ basic \textbar{} full} & \textbf{MOS$\uparrow$ basic \textbar{} full} \\
\cmidrule(lr){3-3} \cmidrule(lr){4-4} \cmidrule(lr){5-5} \cmidrule(lr){6-6}
\multirow{4}{*}{\textbf{Insertion}}
& FluentSpeech   & 12.00 \textbar{} 11.91 & 0.60 \textbar{} 0.60 & 2.90 \textbar{} 2.91 & 3.46 \textbar{} 3.45 \\
& VoiceCraft     & 10.70 \textbar{} 12.94 & 0.67 \textbar{} 0.67 & 3.00 \textbar{} 3.00 & 3.62 \textbar{} 3.60 \\
& Ming-UniAudio$^\diamond$ & 6.63 \textbar{} 7.59 & \textbf{\textcolor{blue}{0.79}} \textbar{} \textbf{\textcolor{blue}{0.79}} & - \textbar{} - & - \textbar{} - \\
& Ours           & \textbf{\textcolor{blue}{4.70}} \textbar{} \textbf{\textcolor{blue}{5.12}} & \textbf{\textcolor{red}{0.82}} \textbar{} \textbf{\textcolor{red}{0.82}} & \textbf{\textcolor{blue}{3.14}} \textbar{} \textbf{\textcolor{blue}{3.13}} & \textbf{\textcolor{blue}{3.86}} \textbar{} \textbf{\textcolor{blue}{3.84}} \\
& Ours (w. GRPO) & \textbf{\textcolor{red}{4.50}} \textbar{} \textbf{\textcolor{red}{4.97}} & \textbf{\textcolor{red}{0.82}} \textbar{} \textbf{\textcolor{red}{0.82}} & \textbf{\textcolor{red}{3.17}} \textbar{} \textbf{\textcolor{red}{3.18}} & \textbf{\textcolor{red}{4.01}} \textbar{} \textbf{\textcolor{red}{3.95}} \\
\midrule

& & \textbf{WER(\%)$\downarrow$ basic \textbar{} full} & \textbf{SIM$\uparrow$ basic \textbar{} full} & \textbf{DNSMOS$\uparrow$ basic \textbar{} full} & \textbf{MOS$\uparrow$ basic \textbar{} full} \\
\cmidrule(lr){3-3} \cmidrule(lr){4-4} \cmidrule(lr){5-5} \cmidrule(lr){6-6}
\multirow{4}{*}{\textbf{Deletion}}
& FluentSpeech   & 8.16 \textbar{} 8.78 & 0.51 \textbar{} 0.52 & 2.91 \textbar{} 2.91 & 3.47 \textbar{} 3.46 \\
& VoiceCraft     & 16.99 \textbar{} 17.88 & 0.60 \textbar{} 0.62 & 3.01 \textbar{} 3.04 & 3.34 \textbar{} 3.36 \\
& Ming-UniAudio$^\diamond$ & 14.85 \textbar{} 27.60 & \textbf{\textcolor{blue}{0.76}} \textbar{} 0.74 & - \textbar{} - & - \textbar{} - \\
& Ours           & \textbf{\textcolor{blue}{7.38}} \textbar{} \textbf{\textcolor{blue}{7.70}} & \textbf{\textcolor{blue}{0.76}} \textbar{} \textbf{\textcolor{blue}{0.77}} & \textbf{\textcolor{blue}{3.07}} \textbar{} \textbf{\textcolor{blue}{3.08}} & \textbf{\textcolor{blue}{3.79}} \textbar{} \textbf{\textcolor{blue}{3.80}} \\
& Ours (w. GRPO) & \textbf{\textcolor{red}{6.91}} \textbar{} \textbf{\textcolor{red}{6.88}} & \textbf{\textcolor{red}{0.77}} \textbar{} \textbf{\textcolor{red}{0.78}} & \textbf{\textcolor{red}{3.09}} \textbar{} \textbf{\textcolor{red}{3.09}} & \textbf{\textcolor{red}{3.88}} \textbar{} \textbf{\textcolor{red}{3.87}} \\
\midrule

& & \textbf{WER(\%)$\downarrow$ basic \textbar{} full} & \textbf{SIM$\uparrow$ basic \textbar{} full} & \textbf{DNSMOS$\uparrow$ basic \textbar{} full} & \textbf{MOS$\uparrow$ basic \textbar{} full} \\
\cmidrule(lr){3-3} \cmidrule(lr){4-4} \cmidrule(lr){5-5} \cmidrule(lr){6-6}
\multirow{4}{*}{\textbf{Substitution}}
& FluentSpeech   & 4.66 \textbar{} 4.65 & 0.51 \textbar{} 0.51 & 2.92 \textbar{} 2.92 & 3.49 \textbar{} 3.48 \\
& VoiceCraft     & 11.98 \textbar{} 12.73 & \textbf{\textcolor{blue}{0.58}} \textbar{} 0.59 & 3.01 \textbar{} 3.02 & 3.57 \textbar{} 3.58 \\
& Ming-UniAudio$^\diamond$ & 8.99 \textbar{} 7.64 & \textbf{\textcolor{red}{0.78}} \textbar{} \textbf{\textcolor{blue}{0.77}} & - \textbar{} - & - \textbar{} - \\
& Ours           & \textbf{\textcolor{blue}{4.40}} \textbar{} \textbf{\textcolor{blue}{4.61}} & \textbf{\textcolor{red}{0.78}} \textbar{} \textbf{\textcolor{red}{0.78}} & \textbf{\textcolor{blue}{3.12}} \textbar{} \textbf{\textcolor{blue}{3.09}} & \textbf{\textcolor{blue}{3.88}} \textbar{} \textbf{\textcolor{blue}{3.86}} \\
& Ours (w. GRPO) & \textbf{\textcolor{red}{4.13}} \textbar{} \textbf{\textcolor{red}{4.41}} & \textbf{\textcolor{red}{0.78}} \textbar{} \textbf{\textcolor{red}{0.78}} & \textbf{\textcolor{red}{3.15}} \textbar{} \textbf{\textcolor{red}{3.11}} & \textbf{\textcolor{red}{3.96}} \textbar{} \textbf{\textcolor{red}{3.93}} \\
\bottomrule
\end{tabular}
} 


\end{table*}

\subsubsection{Gated Self-Consistency Rewards Fusion}

To balance acoustic naturalness ($r_{\text{sc}}$) and content accuracy ($r_{\text{wer}}$), we introduce a gated reward aggregation strategy that filters low-quality samples via a hard validity constraint. The total reward $R$ is defined as:

\begin{equation}
    \small
    R =
    \begin{cases}
    R_{\text{base}} + r_{\text{sc}} + r_{\text{wer}}, & \text{if } \mathbb{I}_{\text{valid}}, \\
    0, & \text{otherwise},
    \end{cases}
\end{equation}
where $R_{\text{base}}$ is a shaping constant ensuring positive rewards for valid samples.

The validity indicator $\mathbb{I}_{\text{valid}}$ enforces both intelligibility and duration consistency:
\begin{equation}
    \small
    \mathbb{I}_{\text{valid}} =
    \underbrace{\left( \operatorname{WER}(\mathbf{W}_{\text{rec}}, \mathbf{T}_{\text{tgt}}) \le \tau_{\text{wer}} \right)}_{\text{Content Integrity}}
    \land
    \underbrace{\left( \left| \frac{L_{\text{gen}} - L_{\text{gt}}}{L_{\text{gt}}} \right| \le \tau_{\text{len}} \right)}_{\text{Duration Stability}},
\end{equation}
where $L_{\text{gen}}=|\hat{\mathbf{S}}_{\text{mid}}|$ and $L_{\text{gt}}=|\mathbf{S}_{\text{mid}}|$. The thresholds $\tau_{\text{wer}}$ and $\tau_{\text{len}}$ specify admissible tolerances. This gating removes invalid samples from optimization, stabilizing GRPO training.

\section{Experiments}
\label{sec:experiments}

\subsection{Experimental Settings}
\textbf{Datasets.}
Training is conducted on \textsf{Libriheavy}~\cite{kang2024libriheavy}, a 50k-hour English speech corpus from LibriVox. 
We evaluate editing performance on two benchmarks. 
First, we adopt the Ming-Freeform-Audio-Edit-Benchmark~\cite{yan2025ming} (basic \& full), covering insertion, deletion, and substitution tasks. Ground-truth editing intervals are obtained via forced alignment using WhisperX~\cite{bain2023whisperx}. 
Second, to assess robustness under varying edit durations, we construct an additional test set from Seed-TTS-Eval~\cite{anastassiou2024seed} by sampling 200 utterances ($>$4s) and applying random masks with durations $\{0.5\text{s}, \dots, 2.5\text{s}\}$.

\noindent\textbf{Baselines.}
We benchmark our framework against three representative models covering diverse editing paradigms:
\begin{itemize}
    \item \textbf{FluentSpeech}~\cite{jiang2023fluentspeech}: A NAR diffusion-based model generating mel-spectrograms directly.
    \item \textbf{VoiceCraft}~\cite{peng2024voicecraft}: A SOTA AR NCLM operating on quantized acoustic tokens.
    \item \textbf{Ming-UniAudio}~\cite{yan2025ming}: A unified LLM designed for speech understanding, editing, and generation.
\end{itemize}

\noindent\textbf{Evaluation Metrics.}
We report both objective and subjective metrics:
\textbf{WER} (Whisper~\cite{radford2023robust}) for intelligibility and boundary consistency;
\textbf{SIM} (WavLM~\cite{chen2022wavlm}) for speaker preservation;
\textbf{DNSMOS}~\cite{reddy2021dnsmos} for perceptual quality; and
\textbf{MOS}, obtained from 15 raters on 90 samples using a 1--5 naturalness scale.
\subsection{Implementation Details}
Our framework consists of supervised pre-training followed by RL alignment.
We adopt the semantic tokenizer, Flow Matching decoder, and HiFiGAN vocoder from CosyVoice3~\cite{du2025cosyvoice}, keeping them frozen.
The semantic-token LLM is trained on \textsf{Libriheavy} with learning rate $1\times10^{-5}$ for up to 10 epochs (gradient accumulation = 2).
During RL training, we use learning rate $1\times10^{-6}$, batch size 4, rollout group size 8, and KL coefficient $\beta=0.01$. Training runs for 400 steps with gradient accumulation 10.
The log-probability reward is computed using CosyVoice3~\cite{du2025cosyvoice}, and the ASR reward using SenseVoiceSmall~\cite{an2024funaudiollm}. Thresholds are set to $\tau_{\text{wer}}=0.2$ and $\tau_{\text{len}}=0.2$.
All experiments are conducted on 8 NVIDIA H800 GPUs.

\subsection{Main Results Analysis}

Table~\ref{tab:main_results_edit} summarizes results on the first benchmark.

\noindent \textbf{Intelligibility and Robustness.}
Our semantic-token-based editing framework achieves the lowest WER across all three editing operations, consistently outperforming SOTA baselines. This advantage arises from decoupling semantic generation from acoustic rendering, which simplifies the modeling objective and reduces prediction instability. Furthermore, incorporating GRPO leads to additional WER reductions across all tasks. We attribute these improvements to the ASR-based reward and length constraint in GRPO, which discourage unintelligible outputs and repetitive generation.

Detailed analysis of different types of editing operations:
\begin{itemize}
    \item \textbf{Insertion:} AR systems generally outperform NAR models, with Ming-UniAudio also showing competitive performance. The NAR baseline (FluentSpeech) exhibits the highest WER, likely due to mask-based prediction, where the predefined masked duration often mismatches the length required by inserted text, resulting in noticeable artifacts.
    \item \textbf{Deletion:} This task proves most challenging for traditional AR models (e.g., VoiceCraft), which suffer from severe hallucinations and often fail to emit the EOS token in time. Conversely, NAR models perform better by simply predicting silence. Our method achieves the best performance; by editing only semantic tokens, we significantly reduce prediction difficulty, enabling precise stopping. Moreover, with GRPO, the WER drops drastically (0.47\% on basic and 0.82\% on full), confirming that the length reward effectively suppresses repetition and incoherent generation.
    \item \textbf{Substitution:} While NAR methods excel here due to similar durations between original and edited text, our method still surpasses the strong NAR baseline. 
\end{itemize}

\begin{table}[t]
\centering
\caption{Robustness evaluation on the subset of Seed-TTS test set across varying masked durations (0.5s to 2.5s). 
\textbf{\textcolor{red}{Red}} indicates the best result, and \textbf{\textcolor{blue}{Blue}} indicates the second best.
}
\label{tab:main_results_edit_duration}
\resizebox{\linewidth}{!}{%
\begin{tabular}{@{}llccccc@{}}
\toprule
\textbf{Metrics} & \textbf{Model} & \multicolumn{5}{c}{\textbf{Masked Duration}} \\
\midrule

& & \textbf{0.5s} & \textbf{1s} & \textbf{1.5s} & \textbf{2s} & \textbf{2.5s} \\
\cmidrule(lr){3-3} \cmidrule(lr){4-4} \cmidrule(lr){5-5} \cmidrule(lr){6-6} \cmidrule(lr){7-7}
\multirow{3}{*}{\textbf{WER(\%) $\downarrow$}}
& FluentSpeech  & 7.533 & 7.107 & 7.231 & 8.300 & 7.390 \\
& VoiceCraft    & 8.504 & 10.505 & 10.813 & 11.525 & 11.190 \\
& Ours          & \textbf{\textcolor{blue}{3.342}} & \textbf{\textcolor{blue}{3.858}} & \textbf{\textcolor{blue}{4.126}} & \textbf{\textcolor{blue}{4.430}} & \textbf{\textcolor{blue}{4.333}} \\
& Ours (w. GRPO) & \textbf{\textcolor{red}{3.202}} & \textbf{\textcolor{red}{3.400}} & \textbf{\textcolor{red}{4.067}} & \textbf{\textcolor{red}{4.117}} & \textbf{\textcolor{red}{4.227}} \\
\midrule

& & \textbf{0.5s} & \textbf{1s} & \textbf{1.5s} & \textbf{2s} & \textbf{2.5s} \\
\cmidrule(lr){3-3} \cmidrule(lr){4-4} \cmidrule(lr){5-5} \cmidrule(lr){6-6} \cmidrule(lr){7-7}
\multirow{3}{*}{\textbf{SIM $\uparrow$}}
& FluentSpeech  & 0.797 & 0.750 & 0.685 & 0.615 & 0.535 \\
& VoiceCraft    & 0.790 & 0.761 & \textbf{\textcolor{blue}{0.723}} & 0.695 & 0.639 \\
& Ours          & \textbf{\textcolor{red}{0.866}} & \textbf{\textcolor{red}{0.855}} & \textbf{\textcolor{red}{0.840}} & \textbf{\textcolor{blue}{0.828}} & \textbf{\textcolor{blue}{0.809}} \\
& Ours (w. GRPO) & \textbf{\textcolor{blue}{0.865}} & \textbf{\textcolor{blue}{0.854}} & \textbf{\textcolor{red}{0.840}} & \textbf{\textcolor{red}{0.829}} & \textbf{\textcolor{red}{0.811}} \\
\midrule

& & \textbf{0.5s} & \textbf{1s} & \textbf{1.5s} & \textbf{2s} & \textbf{2.5s} \\
\cmidrule(lr){3-3} \cmidrule(lr){4-4} \cmidrule(lr){5-5} \cmidrule(lr){6-6} \cmidrule(lr){7-7}
\multirow{3}{*}{\textbf{DNSMOS $\uparrow$}}
& FluentSpeech  & 2.915 & 2.930 & 2.975 & 2.991 & 3.006 \\
& VoiceCraft    & 2.970 & 2.978 & 3.016 & 3.021 & 3.008 \\
& Ours          & \textbf{\textcolor{red}{3.126}} & \textbf{\textcolor{red}{3.139}} & \textbf{\textcolor{blue}{3.138}} & \textbf{\textcolor{blue}{3.138}} & \textbf{\textcolor{blue}{3.128}} \\
& Ours (w. GRPO) & \textbf{\textcolor{blue}{3.124}} & \textbf{\textcolor{blue}{3.138}} & \textbf{\textcolor{red}{3.143}} & \textbf{\textcolor{red}{3.148}} & \textbf{\textcolor{red}{3.148}} \\
\bottomrule
\end{tabular}
} 
\end{table}

\noindent \textbf{Speaker Similarity and Perceptual Quality.}
Our method consistently achieves the highest SIM, DNSMOS, and subjective MOS scores, outperforming all baselines.

In terms of SIM, the Audio-LLM-based Ming-UniAudio scores slightly lower than ours, followed by VoiceCraft, with FluentSpeech performing worst. This validates the superiority of AR-based approaches over diffusion-based NAR methods in capturing speaker characteristics. Notably, GRPO does not significantly alter the SIM score. This is expected, as we optimize the policy for semantic tokens, while timbre preservation is primarily handled by the fixed Flow Matching decoder.

Regarding DNSMOS and Subjective MOS, our method outperforms baselines in naturalness.

\begin{itemize}
    \item Our method significantly outperforms both the AR baseline (VoiceCraft) and the NAR baseline (FluentSpeech).
    \item The introduction of GRPO yields substantial improvements in perceptual metrics. This is driven by the Self-Consistency Rewards, which ensure the style of the edited region aligns seamlessly with the unedited context.
    \item Subjective MOS results mirror the objective DNSMOS, confirming that human listeners perceive our method—especially with GRPO alignment—as the most natural.
\end{itemize}

\subsection{Robustness on Edited Duration}

Table~\ref{tab:main_results_edit_duration} evaluates performance stability as the masked duration increases from 0.5s to 2.5s.

\textbf{Impact on Intelligibility.}
Our method consistently achieves the lowest WER across all durations, with GRPO providing further improvements. Unlike the main benchmark, this setting enforces equal lengths between the masked region and generated content, where the NAR baseline (FluentSpeech) surpasses the traditional AR baseline (VoiceCraft). As edit duration increases, VoiceCraft’s WER rises sharply due to error accumulation in acoustic-token autoregression. In contrast, our method degrades more gradually and remains significantly better than FluentSpeech even at 2.5s, demonstrating strong robustness under long-context editing.

\textbf{Impact on Speaker Similarity.}
Our approach maintains the highest speaker similarity across all durations. Consistent with earlier results, GRPO has minimal influence on SIM, since speaker characteristics are primarily determined by the frozen Flow Matching decoder. Among baselines, VoiceCraft outperforms FluentSpeech, while the NAR model exhibits a pronounced decline as duration increases (dropping to 0.535 at 2.5s). Our method preserves high similarity with only minor decay, benefiting from unified acoustic reconstruction that is largely insensitive to edit length.

\textbf{Impact on Naturalness (DNSMOS).}
Our method consistently achieves the best naturalness scores, with GRPO becoming increasingly beneficial as edit duration grows. Gains are marginal for short edits but widen for longer ones. While baseline methods and our non-aligned model show little improvement or plateau, Ours (with GRPO) shows a clear upward trend. This validates the efficacy of our \textit{Self-Consistency Reward}: by leveraging a pre-trained TTS model to approximate the natural speech distribution, RL optimization guides the model to generate coherent prosody even for complex, long-form edits.

\section{Conclusion}
\label{sec:conclusion}

In this paper, we presented a novel framework for imperceptible text-based speech editing through the principle of \textit{Edit Content, Preserve Acoustics}. 
By shifting the editing operation from the acoustic space to a disentangled semantic space, we established a robust structural foundation for acoustic preservation.
Furthermore, to ensure the edited region fuses indistinguishably with the context, we introduced a Perceptual Alignment stage via Self-Consistency Rewards GRPO. 
Extensive evaluations on two benchmarks demonstrate that our method significantly outperforms SOTA AR and NAR baselines, achieving superior intelligibility, speaker similarity, and perceptual naturalness, even in long-duration scenarios. 
Future work will explore extending this semantic-based framework and the GRPO alignment method to freeform speech editing.

\vfill\pagebreak

\section{Acknowledgments}

This work is supported by the National Natural Science Foundation of China (NSFC) (No. 62322120, No.U2436210, No. 62306316, No. 62206278), and the China Postdoctoral Science Foundation (No. 2025T180461, 2025M771685).



\section{Generative AI Use Disclosure}
We used generative AI tools solely to assist with English writing refinement. These tools were not used to generate or modify the technical content of the paper, including the research ideas, method design, algorithms, mathematical formulations, experimental setup, results, figures, or conclusions, nor to generate citations or attributed text. All authors reviewed, verified, and edited the AI-assisted text, ensured proper attribution and originality, and take full responsibility and accountability for the entire content of the submission.

\bibliographystyle{IEEEtran}
\bibliography{mybib}

\end{document}